\documentclass{article}
\usepackage{spconf,amsmath,amssymb,graphicx}
\usepackage{xcolor}
\usepackage{tikz}
\usetikzlibrary{arrows,positioning,shapes.geometric,decorations.text,decorations.pathmorphing}
\usepackage{pgfplots}
\usepackage{subfig}
\usepackage{multirow}
\usepackage[skip=2pt]{caption}

\usepackage{tabularx}
\usepackage{booktabs}
\newcolumntype{C}[1]{>{\centering\arraybackslash}m{#1}}

\newcommand{\fixme}[2]{\ifx&#2&{\color{red}#1}\else{\color{red}FIXME\{}#1{\color{red}\}}\footnote{{\color{red}#2}}\PackageWarning{Fixme}{#1: #2}\fi}

\def\x{{\mathbf x}}

\title{Partitioned Successive-Cancellation List Decoding of Polar Codes}

\name{\begin{tabular}[x]{@{}c@{}}Seyyed~Ali~Hashemi$^{\star}$, Alexios~Balatsoukas-Stimming$^{\dagger}$, Pascal Giard$^{\star}$,\\Claude Thibeault$^{\diamond}$ and Warren~J.~Gross$^{\star}$\end{tabular}}
\address{$^{\star}$McGill University, Montr\'eal, Qu\'ebec, Canada\\
$^{\dagger}$\'Ecole polytechnique f\'ed\'erale de Lausanne, Lausanne, Switzerland\\
$^{\diamond}$\'Ecole de technologie sup\'erieure, Montr\'eal, Qu\'ebec, Canada}

\usepackage{ifpdf}
\ifpdf
\fi

\begin{document}
\maketitle
\begin{abstract}
Successive-cancellation list (SCL) decoding is an algorithm that provides very good error-correction performance for polar codes. However, its hardware implementation requires a large amount of memory, mainly to store intermediate results. In this paper, a partitioned SCL algorithm is proposed to reduce the large memory requirements of the conventional SCL algorithm. The decoder tree is broken into partitions that are decoded separately. We show that with careful selection of list sizes and number of partitions, the proposed algorithm can outperform conventional SCL while requiring less memory.
\end{abstract}
\section{Introduction}
\label{sec:intro}

Polar codes provably achieve the symmetric capacity of memoryless channels and therefore have gained a lot of attention as promising error-correcting codes~\cite{arikan}. Successive-cancellation (SC) decoding was first proposed as a low-complexity decoding algorithm for polar codes. It was shown that the error probability of polar codes under SC decoding goes to zero as the blocklength goes to infinity, provided that the rate of the polar code is less than the capacity of the channel. From a hardware implementation point of view, SC decoding can be represented as a decoder tree having a fixed time and space complexity and is thus very attractive~\cite{leroux}. However, the algorithm is sub-optimal, especially for decoding moderate-length polar codes.

To improve the error-correction performance of SC decoding, the SC list (SCL) decoding algorithm was proposed in~\cite{tal_list}. Unlike SC decoding, which estimates each bit based on the estimation of previous bits, SCL keeps a constrained list of the most likely candidates at each decoding step using the log-likelihood (LL) of each candidate. SCL reduces the gap between SC and maximum likelihood (ML) decoding at the cost of increased complexity. Furthermore, it was shown that concatenating polar codes with a cyclic redundancy check (CRC) as an outer code improves the performance of SCL to the extent where polar codes decoded with CRC-aided SCL are able to outperform low-density parity-check (LDPC) codes of the same length and rate~\cite{tal_list}. To reduce the hardware complexity associated with LL-based SCL decoding, log-likelihood ratio (LLR) values were used and the path metric calculations adapted accordingly in~\cite{balatsoukas}. Unfortunately, similarly to its LL-based counterpart, LLR-based SCL decoding requires a large memory to store the intermediate values, i.e. the total core area is often largely dominated by memory \cite{balatsoukas}.

In this paper, a \emph{partitioned} SCL (PSCL) decoding algorithm is proposed in order to reduce the memory requirements associated with SCL decoding. More specifically, PSCL decoding performs SCL decoding on partitions of the decoder tree and only one path candidate is transferred from one partition to the next. As a result, memory can be shared between the different partitions of the code, therefore, significantly reducing the overall memory requirements. Without loss of generality, we propose a CRC-aided scheme.

The paper is organized as follows: Section~\ref{sec:PC} offers a brief overview on polar encoding and decoding. Section~\ref{sec:PSCL} describes the proposed PSCL algorithm and compares its error-correction performance with that of conventional SCL decoding. In Section~\ref{sec:results} hardware implementation results are presented showing memory and total area savings of up to 41\% and 42\%, respectively, at similar error-correction performance. Finally, conclusions are drawn in Section~\ref{sec:conclusion}.

\section{Polar Codes}
\label{sec:PC}

A polar code of length $N = 2^n$ which carries $K$ information bits, denoted by $\mathcal{P}(N,K)$, has rate $R \triangleq \frac{K}{N}$ and is constructed by concatenating two polar codes of length $\frac{N}{2}$. Let us consider an input set $\mathbf{u}_0^{N-1} = \{u_0,u_1,\ldots,u_{N-1}\}$ and a coded set $\mathbf{x}_0^{N-1} = \{x_0,x_1,\ldots,x_{N-1}\}$. The recursive concatenation process can be expressed as a modulo-2 matrix multiplication as in
\begin{equation}
\mathbf{x}_0^{N-1} = \mathbf{u}_0^{N-1} \mathbf{G}^{\otimes n}\text{,} \label{eq1}
\end{equation}
where $\mathbf{G}^{\otimes n}$ is the $n$-th Kronecker product of the polarizing matrix $\mathbf{G}=\bigl[\begin{smallmatrix} 1&0\\ 1&1 \end{smallmatrix} \bigr]$.

\begin{figure}
  \centering
  \begin{tikzpicture}[scale=.57, thick]
  \node at (.5,0) {$0$};
  \node at (.5,-1) {$0$};
  \node at (.5,-2) {$0$};
  \node at (.5,-3) {$u_3$};
  \node at (.5,-4) {$0$};
  \node at (.5,-5) {$u_5$};
  \node at (.5,-6) {$u_6$};
  \node at (.5,-7) {$u_7$};

  \foreach \x in {-6,-4,-2,0}
  {
    \draw [->] (1,\x) -- (2,\x);
    \draw (1,\x-1) -- (2.25,\x-1);

    \draw (2.25,\x) circle [radius=.25];
    \draw (2,\x) -- (2.5,\x);
    \draw (2.25,\x-.25) -- (2.25,\x+.25);

    \draw [->] (2.25,\x-1) -- (2.25,\x-.25);

    \fill (2.25,\x-1) circle [radius=.1];
  }

  \foreach \x in {-4,0}
  {
    \draw [->] (2.5,\x) -- (5,\x);
    \draw [->] (2.25,\x-1) -- (3.5,\x-1);

    \draw (5.25,\x) circle [radius=.25];
    \draw (5,\x) -- (5.5,\x);
    \draw (5.25,\x-.25) -- (5.25,\x+.25);

    \draw (3.75,\x-1) circle [radius=.25];
    \draw (3.5,\x-1) -- (4,\x-1);
    \draw (3.75,\x-1-.25) -- (3.75,\x-1+.25);

    \draw [->] (2.25,\x-2) -- (5.25,\x-2) -- (5.25,\x-.25);
    \fill (5.25,\x-2) circle [radius=.1];
    \draw [->] (2,\x-3) -- (3.75,\x-3) -- (3.75,\x-1-.25);
    \fill (3.75,\x-3) circle [radius=.1];
  }

  \draw [->] (5.5,0) -- (11,0);
  \draw [->] (4,-1) -- (9.5,-1);
  \draw [->] (5.25,-2) -- (8,-2);
  \draw [->] (3.75,-3) -- (6.5,-3);

  \foreach \x in {-1,0}
  {
    \draw [->] (5.5+1.5*\x,\x-4) -- (11.25+1.5*\x,\x-4) -- (11.25+1.5*\x,\x-.25);
    \draw [->] (5.25+1.5*\x,\x-6) -- (11.25+1.5*\x-3,\x-6) -- (11.25+1.5*\x-3,\x-2-.25);
  }

  \foreach \x in {-3,...,0}
  {
    \draw (11.25+1.5*\x,\x) circle [radius=.25];
    \draw (11+1.5*\x,\x) -- (11.5+1.5*\x,\x);
    \draw (11.25+1.5*\x,\x-.25) -- (11.25+1.5*\x,\x+.25);

    \fill (11.25+1.5*\x,\x-4) circle [radius=.1];

    \draw [->] (11.5+1.5*\x,\x) -- (12.5,\x);
    \draw [->] (11.25+1.5*\x,\x-4) -- (12.5,\x-4);
  }

  \node at (13,0) {$x_0$};
  \node at (13,-1) {$x_1$};
  \node at (13,-2) {$x_2$};
  \node at (13,-3) {$x_3$};
  \node at (13,-4) {$x_4$};
  \node at (13,-5) {$x_5$};
  \node at (13,-6) {$x_6$};
  \node at (13,-7) {$x_7$};

  \node at (.5,-7.5) {$\underbrace{}$};
  \node at (2.25,-7.5) {$\underbrace{}$};
  \node at (4.5,-7.5) {$\underbrace{\qquad\qquad}$};
  \node at (9,-7.5) {$\underbrace{\qquad\qquad\qquad\qquad\quad}$};
  \node at (-.5,-8) {level};
  \node at (.5,-8) {0};
  \node at (2.25,-8) {1};
  \node at (4.5,-8) {2};
  \node at (9,-8) {3};

\end{tikzpicture}
  \caption{Polar encoding for $\mathcal{P}(8,4)$.}
  \label{fig1}
\end{figure}
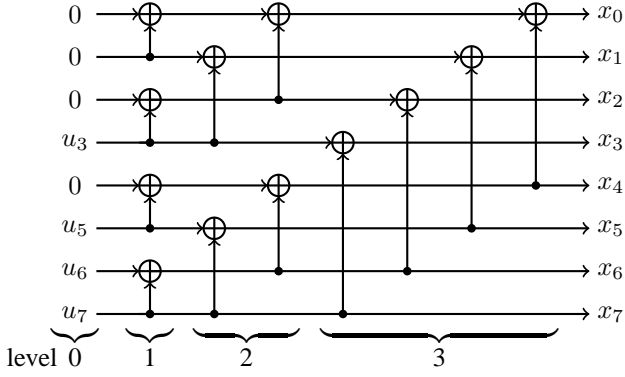

Polar encoding consists of finding the $K$ most reliable bit-channels and transmitting the information bits through them. The $N-K$ least reliable bits are set to a predefined value (usually $0$) which is known by the decoder, and thus are called frozen bits. An example of polar encoding for $\mathcal{P}(8,4)$ is illustrated in Fig.~\ref{fig1}, where $\mathbf{x}_0^{N-1}$ is generated by the encoder before being modulated and sent through the channel. The noisy channel output $\mathbf{y}_0^{N-1}$ is input to the polar decoder.

SC decoding provides each bit estimate $\hat{u}_i$ based on $\mathbf{y}_0^{N-1}$, the previously estimated bits $\mathbf{\hat{u}}_0^{i-1}$, and the location of frozen bits $\mathcal{F}$. The LLR-based formulation is
\begin{equation}
\hat{u}_i =
  \begin{cases}
    0, & \text{if } i \in \mathcal{F} \text{ or } \log\frac{\text{P}(\mathbf{y}_0^{N-1},\mathbf{\hat{u}}_0^{i-1}|\hat{u}_i = 0)}{\text{P}(\mathbf{y}_0^{N-1},\mathbf{\hat{u}}_0^{i-1}|\hat{u}_i = 1)}\geq 0,\\
    1, & \text{otherwise}.
  \end{cases} \label{eq2}
\end{equation}

\begin{figure}
  \centering
  \begin{tikzpicture}[scale=1.9, thick]

  \fill (0,0) circle [radius=.05];

  \fill (-1,-.5) circle [radius=.05];
  \fill (1,-.5) circle [radius=.05];

  \fill (-1.5,-1) circle [radius=.05];
  \fill (-.5,-1) circle [radius=.05];
  \fill (.5,-1) circle [radius=.05];
  \fill (1.5,-1) circle [radius=.05];

  \draw (-1.75,-1.5) circle [radius=.05];
  \draw (-1.25,-1.5) circle [radius=.05];
  \draw (-.75,-1.5) circle [radius=.05];
  \fill (-.25,-1.5) circle [radius=.05];
  \draw (.25,-1.5) circle [radius=.05];
  \fill (.75,-1.5) circle [radius=.05];
  \fill (1.25,-1.5) circle [radius=.05];
  \fill (1.75,-1.5) circle [radius=.05];

  \draw (0,-.05) -- (-1,-.45);
  \draw (0,-.05) -- (1,-.45);

  \draw (-1,-.55) -- (-1.5,-.95);
  \draw (-1,-.55) -- (-.5,-.95);
  \draw (1,-.55) -- (.5,-.95);
  \draw (1,-.55) -- (1.5,-.95);

  \draw (-1.5,-1.05) -- (-1.75,-1.45);
  \draw (-1.5,-1.05) -- (-1.25,-1.45);
  \draw (-.5,-1.05) -- (-.75,-1.45);
  \draw (-.5,-1.05) -- (-.25,-1.45);
  \draw (.5,-1.05) -- (.25,-1.45);
  \draw (.5,-1.05) -- (.75,-1.45);
  \draw (1.5,-1.05) -- (1.25,-1.45);
  \draw (1.5,-1.05) -- (1.75,-1.45);

  \draw [very thin,gray,dashed] (-2,0) -- (2,0);
  \draw [very thin,gray,dashed] (-2,-.5) -- (2,-.5);
  \draw [very thin,gray,dashed] (-2,-1) -- (2,-1);
  \draw [very thin,gray,dashed] (-2,-1.5) -- (2,-1.5);

  \node at (-2.2,.2) {level};
  \node at (-2.2,0) {3};
  \node at (-2.2,-.5) {2};
  \node at (-2.2,-1) {1};
  \node at (-2.2,-1.5) {0};

  \draw [->] (-.12,-.05) -- (-1,-.4) node [above=-.1cm,midway,rotate=25] {$\alpha$};
  \draw [->] (-.88,-.45) -- (0,-.1) node [below=-.1cm,midway,rotate=25] {$\beta$};

  \draw [->] (-1.06,-.55) -- (-1.5,-.9) node [above=-.1cm,midway,rotate=40] {$\alpha_{l}$};
  \draw [->] (-1.44,-.95) -- (-1.0,-0.6) node [below=-.1cm,midway,rotate=40] {$\beta_{l}$};

  \draw [<-] (-.94,-.55) -- (-.5,-.9) node [above=-.1cm,midway,rotate=-40] {$\beta_{r}$};
  \draw [<-] (-.56,-.95) -- (-0.975,-.625) node [below=-.1cm,midway,rotate=-40] {$\alpha_{r}$};

\end{tikzpicture}
  \caption{SC decoder tree for $\mathcal{P}(8,4)$.}
  \label{fig2}
\end{figure}
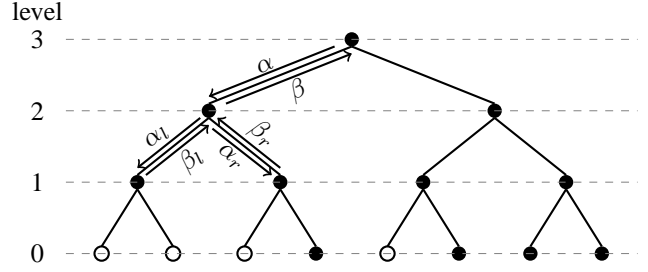

SC works on a decoder tree such as the one illustrated in Fig.~\ref{fig2}. There are two types of messages passed through the different levels in a decoder tree: 1) the soft messages which contain the LLR values $\alpha$; 2) the hard bit estimates $\beta$. Each node at level $s$ of the decoder tree contains $2^s$ bits and the messages in Fig.~\ref{fig2} are calculated as
\begin{align}
\alpha_{l}[i] =& \text{sgn}(\alpha[i])\text{sgn}(\alpha[i+2^{s-1}])\min(|\alpha[i]|,|\alpha[i+2^{s-1}]|)  \text{,} \nonumber\\
\alpha_{r}[i] =& \alpha[i+2^{s-1}] + (1-2\beta_l[i])\alpha[i] \text{,}
\label{eq3}
\end{align}
and
\begin{equation}
\beta[i] =
  \begin{cases}
    \beta_l[i]\oplus \beta_r[i], & \text{if} \quad i < 2^{s-1}\\
    \beta_r[i+2^{s-1}], & \text{otherwise},
  \end{cases}
  \label{eq4}
\end{equation}
where $\oplus$ denotes the bitwise XOR operation \cite{leroux} and $\beta[i]$ are called \emph{partial sums}.

To improve the error-correcting performance of SC decoding, for each non-frozen bit, SCL decoding creates two tentative paths on the decoding tree corresponding to $\hat{u}_i = 0$ and $\hat{u}_i = 1$. In order to avoid an exponential growth in the number of tentative paths, only the $L$ best (i.e., most likely) paths are kept. Specifically, in LLR-based SCL decoding~\cite{balatsoukas}, the $L$ best paths are determined by the following path metric	
\begin{equation}
\text{PM}_i^l =
  \begin{cases}
    \text{PM}_{i-1}^l, & \text{if } \hat{u}_i^l = \frac{1}{2}\left(1-\text{sgn}\left(\alpha_i^l\right)\right),\\
    \text{PM}_{i-1}^l + |\alpha_i^l|, & \text{otherwise},
  \end{cases}
  \label{eq5}
\end{equation}
where $l$ is the path index and $\alpha_i^l$ is the LLR value associated with the $i$-th bit at path $l$. A smaller path metric indicates a more reliable path.

Unfortunately, SCL decoding requires a large amount of memory to store the intermediate values. Let us assume that the LLR and path metric values are quantized with $Q_\alpha$ and $Q_{\text{PM}}$ bits, respectively. The total memory requirements for the storage of the LLR values $\alpha$, the path metrics $\text{PM}$, and the partial sum values $\beta$ in the SC and SCL algorithms are~\cite{leroux} 
\begin{equation}
M_{\text{SC}} = \underbrace{\left(2N-1\right) Q_\alpha}_{\alpha \text{ (LLR values)}} + \underbrace{N-1}_{\beta \text{ (partial sums)}} \text{,} \label{eq6}
\end{equation}
and~\cite{balatsoukas}
\begin{equation}
M_{\text{SCL}} = \underbrace{\left(N+\left(N-1\right)L\right) Q_\alpha}_{\alpha \text{ (LLR values)}} + \underbrace{LQ_\text{PM}}_\text{path metrics} + \underbrace{\left(2N-1\right)L}_{\beta \text{ (partial sums)}}, \label{eq7}
\end{equation}
respectively. We note that $Q_{\text{PM}}$ grows at most as $\log N$ \cite{balatsoukas}, so the term $LQ_\text{PM}$ is negligible in Eq.~\eqref{eq7}.

\section{Partitioned SCL Decoding of\\ Polar Codes}
\label{sec:PSCL}

The large memory requirements of the SCL algorithm translate into a large area occupation in the actual hardware decoder implementation. In fact, the total area is often largely dominated by memory, e.g. the memory area accounts for 45\% of the total area in \cite{balatsoukas}. In order to reduce the required memory and, therefore, the area of the decoder, we propose a \emph{partitioned} SCL (PSCL) decoding technique. 

\subsection{Proposed PSCL Decoding Algorithm}
The conventional CRC-aided SCL decoding algorithm first performs SCL decoding to obtain the $L$ most likely codeword candidates and, in the end, selects the (hopefully) correct estimate by choosing the candidate that matches the expected CRC. If no codeword verifies the CRC, the candidate with the best path metric is selected. 

\begin{figure}
  \centering
  \begin{tikzpicture}[scale=1.9, thick]

  \fill (0,0) circle [radius=.05];

  \fill (-1,-.5) circle [radius=.05];
  \fill (1,-.5) circle [radius=.05];

  \draw (0,-.05) -- (-1,-.45);
  \draw (0,-.05) -- (1,-.45);

  \draw [very thin,gray,dashed] (-2,0) -- (2,0);
  \draw [very thin,gray,dashed] (-2,-.5) -- (2,-.5);

  \node at (-2.25,.2) {level};
  \node at (-2.25,0) {$n$};
  \node at (-2.25,-.5) {$n-1$};

  \draw [->] (-.12,-.05) -- (-1,-.4) node [above=-.1cm,midway,rotate=25] {$\alpha_l$};
  \draw [->] (-.88,-.45) -- (0,-.1) node [below=-.1cm,midway,rotate=25] {$\beta_l$};

  \draw [<-] (.12,-.05) -- (1,-.4) node [above=-.1cm,midway,rotate=-25] {$\beta_r$};
  \draw [<-] (.88,-.45) -- (0.1,-.15) node [below=-.1cm,midway,rotate=-25] {$\alpha_r$};

  \draw (-1.5,-.55) rectangle (-.5,-1.1) node[midway,align=center,text width=1.7cm] {CRC-aided SCL};

  \draw (.5,-.55) rectangle (1.5,-1.1) node[midway,align=center,text width=1.7cm] {CRC-aided SCL};

\end{tikzpicture}
  \caption{PSCL with two partitions.}
  \label{fig3}
\end{figure}
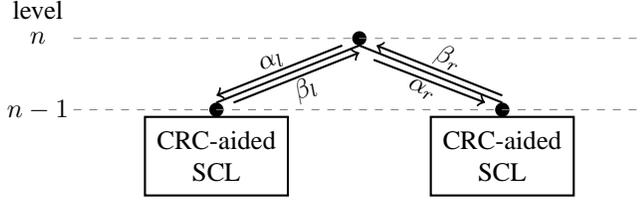

\begin{figure}
  \centering
  \begin{tikzpicture}[scale=1.9, thick]

  \fill (0,0) circle [radius=.05];

  \fill (-1,-.5) circle [radius=.05];
  \fill (1,-.5) circle [radius=.05];

  \fill (-1.5,-1) circle [radius=.05];
  \fill (-.5,-1) circle [radius=.05];
  \fill (.5,-1) circle [radius=.05];
  \fill (1.5,-1) circle [radius=.05];

  \draw (0,-.05) -- (-1,-.45);
  \draw (0,-.05) -- (1,-.45);

  \draw (-1,-.55) -- (-1.5,-.95);
  \draw (-1,-.55) -- (-.5,-.95);
  \draw (1,-.55) -- (.5,-.95);
  \draw (1,-.55) -- (1.5,-.95);

  \draw [very thin,gray,dashed] (-2,0) -- (2,0);
  \draw [very thin,gray,dashed] (-2,-.5) -- (2,-.5);
  \draw [very thin,gray,dashed] (-2,-1) -- (2,-1);

  \node at (-2.25,.2) {level};
  \node at (-2.25,0) {$n$};
  \node at (-2.25,-.5) {$n-1$};
  \node at (-2.25,-1) {$n-2$};

  \draw [->] (-.12,-.05) -- (-1,-.4) node [above=-.1cm,midway,rotate=25] {$\alpha$};
  \draw [->] (-.88,-.45) -- (0,-.1) node [below=-.1cm,midway,rotate=25] {$\beta$};

  \draw [->] (-1.06,-.55) -- (-1.5,-.9) node [above=-.1cm,midway,rotate=40] {$\alpha_{l}$};
  \draw [->] (-1.44,-.95) -- (-1,-.6) node [below=-.1cm,midway,rotate=40] {$\beta_{l}$};

  \draw [<-] (-.94,-.55) -- (-.5,-.9) node [above=-.1cm,midway,rotate=-40] {$\beta_{r}$};
  \draw [<-] (-.56,-.95) -- (-0.95,-.65) node [below=-.1cm,midway,rotate=-40] {$\alpha_{r}$};

  \draw (-1.95,-1.05) rectangle (-1.05,-1.6) node[midway,align=center,text width=1.7cm] {CRC-aided SCL};

  \draw (-.95,-1.05) rectangle (-.05,-1.6) node[midway,align=center,text width=1.7cm] {CRC-aided SCL};

  \draw (.05,-1.05) rectangle (.95,-1.6) node[midway,align=center,text width=1.7cm] {CRC-aided SCL};

  \draw (1.05,-1.05) rectangle (1.95,-1.6) node[midway,align=center,text width=1.7cm] {CRC-aided SCL};

\end{tikzpicture}
  \caption{PSCL with four partitions.}
  \label{fig4}
\end{figure}
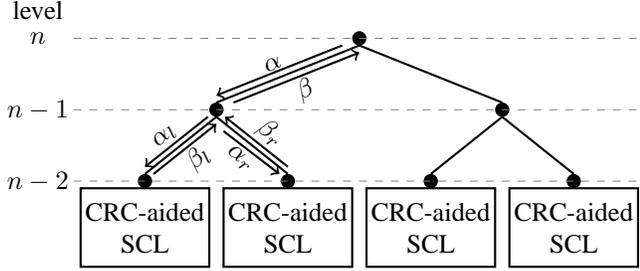

In PSCL decoding, on the other hand, the decoder tree is broken into partitions (i.e., subtrees) and SCL decoding is performed only on the partitions, while the standard SC rules are applied to the remainder of the decoding tree. Each partition outputs a single candidate codeword which is selected with the help of a CRC and then sent to the next partition for further decoding. The decoding process starts with the standard SC update rules given by~\eqref{eq3} and~\eqref{eq4}. Therefore, the decoder does not require memory to store $L$ entire trees of internal LLRs, but only $L$ copies of the partitions on which SCL decoding is performed.  

Fig.~\ref{fig3} and Fig.~\ref{fig4} show the PSCL process when a code is broken into two and four partitions, respectively. The memory used in each CRC-aided SCL decoding block can be shared with the next decoding block since only one candidate survives after decoding each partition. The total memory usage in PSCL with $P$ partitions and list size $L$ can be calculated as
\begin{align}
	M_{\text{PSCL}} = & \underbrace{\left(\sum_{k = 0}^{P-1}\frac{N}{2^k} + \left(\frac{N}{2^{P-1}}-1\right)L\right)Q_{\alpha}}_{\alpha \text{ (LLR values)}} + \underbrace{LQ_{\text{PM}}}_{\text{path metrics}}\nonumber\\ &+ \underbrace{\sum_{k = 1}^{P-2}\frac{N}{2^k} + \left(\frac{N}{2^{P-2}}-1\right)L}_{\beta \text{ (partial sums)}}, \label{eq8}
\end{align}
where $P\geq 2$ and $P=1$ makes PSCL decoding equivalent to conventional SCL decoding. Also note that when $P=2$, $\sum_{k = 1}^{P-2}\frac{N}{2^k}=0$.

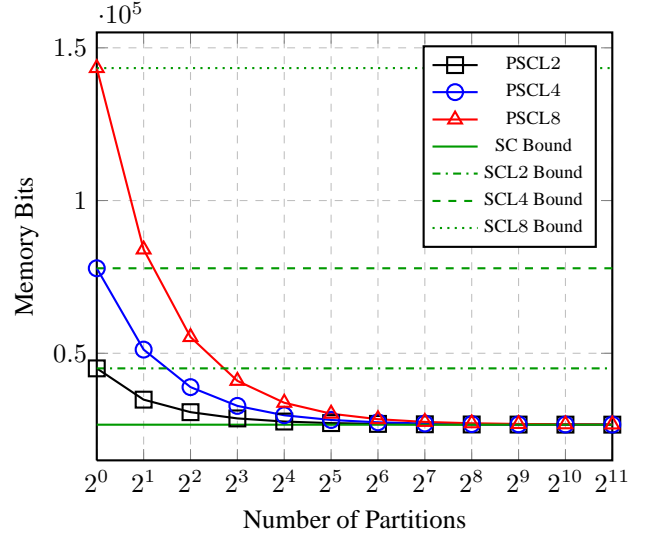
\begin{figure}
  \centering
  \begin{tikzpicture}

\begin{axis}[
	scale = 1,
    xmode=log,
    log basis x=2,
    xtick=data,
    xlabel={Number of Partitions},
    ylabel={Memory Bits}, ylabel style={yshift=-0.85em},
    legend pos=north east,
    grid=both,
    ymajorgrids=true,
    xmajorgrids=true,
    grid style=dashed,
    xmin = 1,
    xmax = 2048,
    thick,
    mark size = 3,
    legend style={
      font=\fontsize{7pt}{7.2}\selectfont,
    },
]

\addplot[
    color=black,
    mark=square,
]
table {
1 45058
2 34818
4 30722
8 28674
16 27650
32 27138
64 26882
128 26754
256 26690
512 26658
1024 26642
2048 26634
};
\addlegendentry{PSCL$2$}

\addplot[
    color=blue,
    mark=o,
]
table {
1 77828
2 51204
4 38916
8 32772
16 29700
32 28164
64 27396
128 27012
256 26820
512 26724
1024 26676
2048 26652
};
\addlegendentry{PSCL$4$}

\addplot[
    color=red,
    mark=triangle,
]
table {
1 143368
2 83976
4 55304
8 40968
16 33800
32 30216
64 28424
128 27528
256 27080
512 26856
1024 26744
2048 26688
};
\addlegendentry{PSCL$8$}

\addplot[
    color=green!60!black,
]
table {
1 26617
2048 26617
};
\addlegendentry{SC Bound}

\addplot[
    color=green!60!black,
    dashdotted,
]
table {
1 45058
2048 45058
};
\addlegendentry{SCL$2$ Bound}

\addplot[
    color=green!60!black,
    dashed,
]
table {
1 77828
2048 77828
};
\addlegendentry{SCL$4$ Bound}

\addplot[
    color=green!60!black,
    dotted,
]
table {
1 143368
2048 143368
};
\addlegendentry{SCL$8$ Bound}

\end{axis}
\end{tikzpicture}
  \caption{Memory requirements for polar codes of length $N=2048$. PSCL$L$ (SCL$L$) denotes the PSCL (SCL) decoding algorithm with list size $L$.}
  \label{fig5}
\end{figure}

It should be noted that the lower bound on the memory usage for PSCL is the memory requirement of the SC algorithm and the upper bound is the memory required by SCL with list size $L$. Fig.~\ref{fig5} illustrates the PSCL memory usage with different numbers of partitions and list sizes for a polar code with $N = 2048$, $Q_\alpha = 6$ bits, and $Q_\text{PM} = 8$ bits. As it can be seen in the figure, the amount of memory decays exponentially towards the SC bound as the number of partitions increases.
In other words, a small increase in the number of partitions results in significant savings, e.g. using four partitions in PSCL$4$ is expected to require less memory than SCL$2$.

\subsection{Error-Correction Performance} 
\begin{figure}[t]
  \centering
  \hspace{-15pt}
  \begin{tikzpicture}
  \pgfplotsset{
    label style = {font=\fontsize{9pt}{7.2}\selectfont},
    tick label style = {font=\fontsize{7pt}{7.2}\selectfont}
  }

\begin{axis}[
	scale = 1,
    ymode=log,
    xlabel={$E_b/N_0$ [\text{dB}]}, xlabel style={yshift=0.8em},
    ylabel={FER}, ylabel style={yshift=-0.75em},
    grid=both, ymax=1.3, ymin=6e-7,
    ymajorgrids=true,
    xmajorgrids=true,
    grid style=dashed,
    width=0.5\columnwidth, height=7cm,
    thick,
    mark size=3,
    legend style={
      anchor={center},
      cells={anchor=west},
      column sep= 2mm,
      font=\fontsize{7pt}{7.2}\selectfont,
    },
    legend to name=perf-legend,
    legend columns=2,
]

\addplot[
    color=brown,
    mark=star,
]
table {
1 0.781250000000000
1.5 0.297973778307509
2 0.044748735848212
2.5 0.003519032688295
3 0.000238058622412
};
\addlegendentry{SC}

\addplot[
    color=red,
    mark=o,
]
table {
1 0.730034982508746
1.5 0.189105447276362
2 0.015892053973014
2.5 0.000637586501827
3 0.000024952994796
};
\addlegendentry{PSCL$(2,2)$-CRC$16$}

\addplot[
    color=blue,
    mark=square,
]
table {
1 0.731934032983508
1.5 0.200199900049975
2 0.016391804097951
2.5 0.000518363009616
3 0.000026699704432
};
\addlegendentry{SCL$2$-CRC$32$}

\addplot[
    color=green,
    mark=triangle,
]
table {
1 0.742315369261477
1.5 0.200998003992016
2 0.017065868263473
2.5 0.000875810124365
3 0.000037122005182
};
\addlegendentry{PSCL$(4,2)$-CRC$8$}

\addplot[
    color=orange,
    mark=asterisk,
]
table {
1 0.548226969805008
1.5 0.079230446052746
2 0.002973985261054
2.5 0.000057398699350
3 0.000001060530265
};
\addlegendentry{SCL$4$-CRC$32$}

\addplot[
    color=black,
    mark=diamond,
]
table {
1 0.539100000000000
1.5 0.082600000000000
2 0.003878975950349
2.5 0.000100134179801
3 0.000005134877836
};
\addlegendentry{PSCL$(4,4)$-CRC$8$}
\end{axis}
\end{tikzpicture}
  \begin{tikzpicture}
  \pgfplotsset{
    label style = {font=\fontsize{9pt}{7.2}\selectfont},
    tick label style = {font=\fontsize{7pt}{7.2}\selectfont},
  }

\begin{axis}[
	scale = 1,
    ymode=log,
    xlabel={$E_b/N_0$ [\text{dB}]}, xlabel style={yshift=0.8em},
    ylabel={BER}, ylabel style={yshift=-0.75em},%
    grid=both, ymax=5e-1, ymin=3e-8,
    ymajorgrids=true,
    xmajorgrids=true,
    width=0.5\columnwidth, height=7.0cm,
    grid style=dashed,
    thick,
    mark size=3,
]

\addplot[
    color=brown,
    mark=star,
]
table {
1 0.280780792236328
1.5 0.081545878650179
2 0.008935981689377
2.5 0.000449288374612
3 0.000017683524525
};

\addplot[
    color=red,
    mark=o,
]
table {
1 0.260639309251624
1.5 0.049406937156422
2 0.002951356353073
2.5 0.000086517013646
3 0.000001548352822
};

\addplot[
    color=blue,
    mark=square,
]
table {
1 0.258614540386057
1.5 0.051252108320840
2 0.002869268490755
2.5 0.000061221506233
3 0.000001864651525
};

\addplot[
    color=green,
    mark=triangle,
]
table {
1 0.274232492047156
1.5 0.055536680545160
2 0.003069349582086
2.5 0.000110699320700
3 0.000002067449175
};

\addplot[
    color=orange,
    mark=asterisk,
]
table {
1 0.182716617467769
1.5 0.019235253002474
2 0.000489478748444
2.5 0.000005824550533
3 0.000000059641563
};

\addplot[
    color=black,
    mark=diamond,
]
table {
1 0.192196484375000
1.5 0.023565917968750
2 0.000818259370151
2.5 0.000013698042877
3 0.000000272840530
};
\end{axis}
\end{tikzpicture}
  \\
  \hspace{20pt}\ref{perf-legend}\vspace{2pt}
  \caption{Frame error rate (FER) and bit error rate (BER) performance comparison between CRC-aided SCL and PSCL decoding of $\mathcal{P}(2048,1024)$. The code is optimized for $E_b/N_0 = 2~\text{dB}$.}
  \label{fig6}
\end{figure}
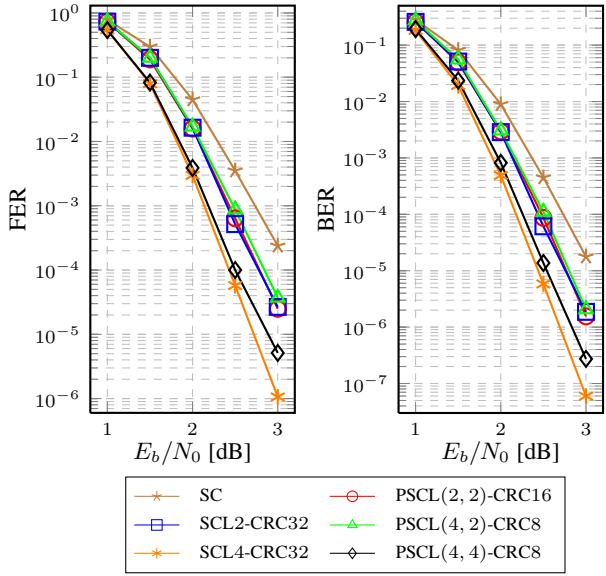

Fig.~\ref{fig6} shows the frame error rate (FER) and bit error rate (BER) performance of SCL and PSCL. The error-correction performance of the plain SC algorithm is also included as a reference. SCL$L$-CRC$x$ denotes the SCL algorithm with list size $L$ and CRC length $x$ and PSCL$(P,L)$-CRC$x$ represents the PSCL algorithm with $P$ partitions, list size $L$, and a CRC of length $x$.

The performance results are provided for a polar code of length $N=2048$ and rate $R = \frac{1}{2}$, while a CRC of length $32$ is used for the conventional SCL decoding algorithm. To keep the code rate unchanged and to have a fair comparison, PSCL$(2,L)$ uses a CRC of length $16$, i.e. each of its two partitions uses a CRC$16$. Similarly, for PSCL$(4,L)$ each of the four partitions uses a CRC$8$. The CRC polynomials were taken from \cite{koopman2004cyclic,koopman200232}.

Fig.~\ref{fig6} shows that PSCL$(2,2)$-CRC$16$ has identical FER and BER performance compared to SCL$2$-CRC$32$ and there is only a slight deterioration in performance when the code is broken into four partitions, as shown by the PSCL$(4,2)$-CRC$8$ curve. However, in Fig.~\ref{fig6}, it can also be seen that PSCL$(4,4)$-CRC$8$ has superior error-correction performance compared to that of SCL$2$-CRC$32$. Furthermore, PSCL$(4,4)$ actually requires slightly less memory than SCL$2$, as shown in Fig.~\ref{fig5}. Thus, PSCL achieves better performance \emph{and} reduces memory usage at the same time.

\section{Hardware Implementation Results}\label{sec:results}
Table~\ref{tab:synthesis} presents indicative synthesis results to compare SC, the conventional CRC-aided SCL algorithm with $L \in \{2,4\}$, and the proposed PSCL algorithm for $L \in \{2,4\}$ and $P \in \{2,4\}$, for a polar code of blocklength $N = 2048$. For the CRC-aided SCL algorithm, the hardware architecture of~\cite{balatsoukas} is used while an appropriately modified version of \cite{balatsoukas} was used for the PSCL algorithm. All synthesis results are for a TSMC $90$~nm CMOS library ($1$~V, $25^\circ$C) with a target frequency of $500$ MHz. All decoders have an equal latency of $5248$ clock cycles ($10.5$~$\mu$s) and throughput of $164$ Mbps.

From Table~\ref{tab:synthesis}, we observe that the PSCL$(2,2)$ and PSCL$(4,2)$ decoders require $23\%$ and $41\%$ less memory area than the SCL$2$ decoder, respectively. The PSCL$(4,4)$ decoder implementation is shown to require $23\%$ less memory area than the SCL$2$ decoder while offering a better coding gain by approximately $0.25$ dB at a BER of $10^{-5}$. Regardless of the implementation, the memory area of the list decoders amounts to $40\%$--$45\%$ of the total area. The memory savings observed for the PSCL implementations thus translate into very significant reductions in the total area, making them very attractive compared to the conventional SCL decoders.

\begin{table}[t]
	\centering
	\caption{Synthesis area results for the SC, CRC-aided SCL, and PSCL decoding algorithms.}\label{tab:synthesis}
  \small
	\begin{tabular}{lcc}
    \toprule
    Algorithm   				& Total (mm$^2$) & Memory (mm$^2$) \\
    \midrule
		SC									& $0.723$								& $0.413$\\
    SCL$2$-CRC$32$      & $1.563$               & $0.702$\\
    SCL$4$-CRC$32$      & $3.075$               & $1.214$\\
    PSCL$(2,2)$-CRC$16$ & $1.189$               & $0.540$\\
    PSCL$(4,2)$-CRC$8$  & $0.909$               & $0.415$\\
    PSCL$(4,4)$-CRC$8$  & $1.356$               & $0.543$\\
    \bottomrule
	\end{tabular}
\end{table}

\section{Conclusion}
\label{sec:conclusion}
In this paper, we have proposed a novel partitioned list decoding algorithm for polar codes. In this algorithm, the code is broken into partitions and each partition is decoded with a CRC-aided successive-cancellation list decoder. Since the memory is shared between different partitions in the code, the memory requirements of a hardware implementation of partitioned list decoder is significantly smaller than that of a conventional list decoder without any error-correction performance degradation. Implementation results show that at equivalent error-correction performance, the proposed algorithm leads to memory and total area savings of $41\%$ and $42\%$, respectively, when compared to a similar list decoder implementation. Moreover, the proposed algorithm enables a coding gain of approximately $0.25$ dB at a bit error rate of $10^{-5}$ while occupying $13\%$ less total area than the conventional CRC-aided successive-cancellation list decoder.

\section*{ACKNOWLEDGEMENT}
The authors would like to thank Gabi~Sarkis and Carlo~Condo of McGill University for helpful discussions.

\bibliographystyle{IEEEbib}
\bibliography{IEEEabrv,refs}

\end{document}